\documentclass[11pt]{article}

\usepackage[T1]{fontenc}
\usepackage[utf8]{inputenc}
\usepackage{lmodern}
\usepackage{microtype}
\usepackage[margin=1in]{geometry}
\usepackage{authblk}
\usepackage{graphicx}
\usepackage{tabularx}
\usepackage[table]{xcolor}
\usepackage{array}
\usepackage{amssymb}
\usepackage[hidelinks]{hyperref}
\usepackage[nolist]{acronym}
\usepackage[normalem]{ulem}
\usepackage{xparse}
\usepackage{makecell}
\usepackage{tikz}
\usepackage{xurl}
\usepackage{orcidlink}
\usetikzlibrary{positioning, arrows.meta, shapes.geometric, fit, backgrounds}

\title{Quantum--HPC Software Stacks and the openQSE Reference Architecture: A Survey}
\date{}

\author[1]{Amir Shehata\thanks{Corresponding author: shehataa@ornl.gov}\,\orcidlink{0000-0002-2453-1426}}
\author[2]{Brian Austin\,\orcidlink{0009-0005-5881-1927}}
\author[1]{Tom Beck\,\orcidlink{0000-0001-8973-7145}}
\author[3,4]{Lukas Burgholzer\,\orcidlink{0000-0003-4699-1316}}
\author[5]{Alex Chernoguzov\,\orcidlink{0000-0003-4034-6850}}
\author[6]{Spencer Churchill\,\orcidlink{0000-0001-9755-236X}}
\author[7]{Andrea Delgado\,\orcidlink{0000-0003-3453-7204}}
\author[8]{Yasuko Eckert}
\author[9]{Jeffery Heckey}
\author[10]{Kevin Kissell}
\author[2]{Katherine Klymko\,\orcidlink{0000-0002-4158-5776}}
\author[6]{Josh Moles\,\orcidlink{0009-0009-1575-3463}}
\author[1]{Thomas Naughton\,\orcidlink{0000-0002-3546-2382}}
\author[11]{Lee James O'Riordan\,\orcidlink{0000-0002-6758-9433}}
\author[12]{Christian Ortiz Pauyac\,\orcidlink{0000-0001-5074-8920}}
\author[6]{Guen Prawiroatmodjo\,\orcidlink{0000-0002-3006-2064}}
\author[2]{Ermal Rrapaj\,\orcidlink{0000-0002-3222-7010}}
\author[17]{Emre Sahin\, \orcidlink{0000-0002-5996-0407}}
\author[6]{Jiri Schindler\,\orcidlink{0000-0002-3592-6634}}
\author[13]{Laura Schulz\,\orcidlink{0000-0002-4702-3440}}
\author[9]{Sebastian Stern}
\author[9]{Tyler Takeshita\,\orcidlink{0000-0003-0067-2846}}
\author[14,15]{Miwako Tsuji\,\orcidlink{0000-0003-4709-1969}}
\author[17]{Oscar Wallis\, \orcidlink{0009-0002-7323-2059}}
\author[16]{Aleksander Wennersteen\,\orcidlink{0009-0006-5486-0980}}
\author[1]{Travis Humble\,\orcidlink{0000-0002-9449-0498}}
\author[4]{Martin Schulz\,\orcidlink{0000-0001-9013-435X}}

\affil[1]{Oak Ridge National Laboratory, Oak Ridge, TN, USA}
\affil[2]{Lawrence Berkeley National Laboratory, Berkeley, CA, USA}
\affil[3]{Munich Quantum Software Company, Munich, Germany}
\affil[4]{Technical University of Munich, Munich, Germany}
\affil[5]{Quantinuum, Broomfield, CO, USA}
\affil[6]{IonQ, College Park, MD, USA}
\affil[7]{Qblox, Oak Ridge, TN, USA}
\affil[8]{AMD, Bellevue, WA, USA}
\affil[9]{Amazon Web Services, Seattle, WA, USA}
\affil[10]{Alice and Bob, Paris, France}
\affil[11]{Xanadu Quantum Technologies Inc., Toronto, Canada}
\affil[12]{Quantum Brilliance, Munich, Germany}
\affil[13]{Argonne National Laboratory, Chicago, IL, USA}
\affil[14]{RIKEN R-CCS, Kobe, Japan}
\affil[15]{University of Tsukuba, Tsukuba, Japan}
\affil[16]{Pasqal, Palaiseau, France}
\affil[17]{The Hartree Centre, STFC, Warrington, UK}

\begin{document}

\begin{acronym}
\acro{ATD}{application time domain}
\acro{AWS}{Amazon Web Services}
\acro{DTD}{deterministic timing domain}
\acro{FTEE}{fault-tolerant execution engine}
\acro{FTQC}{fault-tolerant quantum computing}
\acro{GRES}{generic resource}
\acro{HAPI}{hardware-provider API}
\acro{HECC}{hierarchical error correction code}
\acro{HPC}{high-performance computing}
\acrodefplural{HPC}[HPC]{high-performance computer}
\acro{IR}{intermediate representation}
\acro{ISA}{instruction set architecture}
\acro{JWT}{JSON web tokens}
\acro{LCL}{logical control layer}
\acro{MLIR}{multi-level intermediate representation}
\acro{MQT}{Munich Quantum Toolkit}
\acro{NISQ}{noisy intermediate-scale quantum}
\acro{openQSE}{open quantum-HPC software ecosystem}
\acro{PBS}{portable batch system}
\acro{PCS}{parallel computing service}
\acro{PTD}{physical timing domain}
\acro{QCSC}{Quantum-Centric Supercomputing}
\acro{QDK}{quantum development kit}
\acro{QDMI}{quantum device management interface}
\acro{QEC}{quantum error correction}
\acro{QHPC}{quantum high-performance computing}
\acro{QIR}{quantum intermediate representation}
\acro{QoS}{quality of service}
\acro{QPU}{quantum processing unit}
\acro{QRI}{quantum runtime interface}
\acro{QRMI}{quantum resource management interface}
\acro{QTIL}{queue utility}
\acro{RMS}{resource management system}
\acro{RTD}{real-time domain}
\acro{Slurm}{simple Linux utility for resource management}
\acro{SPANK}{Slurm plug-in architecture for node and job kontrol}
\acro{SQC}{supercomputer quantum computer computation}
\acro{VQE}{variational quantum eigensolver}

\end{acronym}

\maketitle

\vspace{-5mm}
\begin{abstract}
Quantum resources are increasingly integrated into \ac{HPC} and cloud
environments, but \ac{QHPC} software stacks remain isolated, often
proprietary, full-stack solutions lacking common interfaces across runtime,
resource management, orchestration, and execution layers. This paper analyzes
nine production \ac{QHPC} stacks and identifies common design patterns and
emerging requirements, covering deployment models, application interaction
patterns, SDK support, and readiness for fault-tolerant operation. The survey
exposes consistent needs in runtime abstraction, resource management,
interconnect semantics, and observability. Based on these findings, we propose
the \ac{openQSE} reference architecture as a first step toward unifying the
state-of-the-practice. \ac{openQSE} defines a set of layer boundaries that
allow different implementations to interoperate while preserving deployment
flexibility, and is structured to support both current \ac{NISQ} workloads
and future \ac{FTQC} systems without changes to upper-layer application
interfaces.
\end{abstract}

\acresetall
\section{Overview}

Quantum computers do not operate in isolation. Practical quantum execution
depends on classical software and hardware for compilation, orchestration,
control, scheduling, data movement, and error-management functions. As a
result, integrating quantum resources into \ac{HPC} environments is an
end-to-end systems problem, not only a device-access problem. \ac{HPC} centers
are already deploying and evaluating these systems across multiple providers
and deployment models~\cite{qrmi-slurm, openqse-workshops, Mansfield_Int, iqm_eqxa_2025}.

While large strides have been made in this area over the last few years, it has led
to a situation where
this integration currently occurs through many independent, fragmented software
paths. Many interfaces remain provider-specific, assumptions vary between cloud
and on-premises environments, and sites often rebuild similar
integration logic for job submission, resource description, runtime interaction,
and security handling. This fragmentation increases maintenance cost, slows comparative
evaluations, and weakens portability for applications and operations teams.

In this paper we review the current state-of-the-practice in this area, analyzing the most prominent examples of \ac{QHPC} software stacks. We compare their architectures and design decisions, review key interfaces and user accesss modes, and characterize their current state. Based on that, we gather a comprehensive picture of the current \ac{QHPC} approaches and offer a first-of-a-kind mapping into a single blueprint, which can then be used to drive a common community architecture. This forms the input for the \ac{openQSE} initiative recently formed by major contributors, including \ac{HPC} centers, vendors, and users of \ac{QHPC} systems.

Within that context, this paper analyzes nine representative
\ac{QHPC} stacks, developed by industry and research. They span all major deployment options, i.e., managed cloud services,
scheduler-centric datacenter integration, and brokered-offload designs.
In particular,
we include the \ac{QHPC} stacks from
AWS, IBM, IonQ,
JHPC-Quantum, MQSS, Pasqal, Quantinuum, Quantum Brilliance, and Xanadu, and compare them
based on their key properties.
We explore
convergence points and areas for future development in runtime boundaries,
resource management, and interconnect/control assumptions. We then form a common
reference architecture that is powerful enough to map all identified properties and
mechanisms into a single architecture blueprint.

In summary, our paper makes three main
contributions:
\begin{itemize}
    \item {\bf Quantum-HPC Stack Analysis}: We review the state-of-the-practice in \ac{QHPC} software stacks by examining nine widely used representative frameworks. We identify common architectural patterns and important differences across current approaches and synthesize them into an interface convergence matrix (\autoref{tab:stack-convergence}) that maps where the field is converging and where it is not.

    \item {\bf Quantum-HPC Requirements Framework}: We introduce a
    requirements framework that spans deployment environments,
    hardware modalities, application patterns, and quantum-computing
 eras, providing a structured design space to consider
 current and future \ac{QHPC} software systems.

\item {\bf openQSE Reference Architecture}: We propose the community-driven \ac{openQSE} architecture as a vendor-neutral layered reference architecture that provides a holistic view of the integrated \ac{QHPC} system without assigning ownership of specific layers to vendors or HPC centers. It preserves implementation flexibility, while stabilizing interfaces for portable, interoperable, and future-ready integration.

\end{itemize}

\section{Problem Statement}
Quantum computers are expected to become useful in hybrid scientific workflows
for problem classes, such as Hamiltonian simulation, optimization, and quantum
machine learning, with broader application areas including electronic
structure and related chemistry, materials, and data-analysis workloads~\cite{camps2025quantum, ibm-reference-arch}. In order for those workflows to become
practical for scientists and facilities, though, quantum computers must be integrated
with \ac{HPC} resources in an intelligent and efficient manner. This integration
will avoid repeatedly adapting to different provider-specific interfaces,
runtime assumptions, and deployment models. Furthermore, it will remove operational overhead
for data centers that need to compare evolving quantum platforms on a common
basis and for scientists who need to evaluate multiple technologies.

Many groups and projects are working on this
kind of integration, creating a broad field of individually grown solutions. As we
mature the field, it will be critical to track the state-of-the-practice of current
approaches, to gather differences and similarities, and to track the specific
needs and requirements that drove individual design decisions. For such
a state-of-the-practice survey,
we must first establish a needed framework to understand the fundamentally
different scaling models and allocation semantics that drive the use of
quantum resources.
HPC capacity scales by adding comparatively modular and independently scalable resources such as racks, nodes, cores, memory, and GPUs, which can be allocated incrementally or fractionally to users.
By contrast, the capacity of \acp{QPU} scales through the addition and improvement of qubits, which remain scarce and tightly coupled. They cannot be scheduled efficiently as fractional resources. Quantum information cannot be copied, meaning that they also cannot be pre-empively scheduled or checkpointed.
These distinctions are architecturally significant, as they shape scheduling, fairness, utilization, and the design of hybrid workflows.

\section{Classification Axes}
\label{sec:tec-req}
Any practical \ac{QHPC} architecture must, therefore, account for both sides of the scaling asymmetry as well as handle a broad range of requirements in order to use classical and quantum resources efficiently. To enable a fair comparison in our analysis of the state-of-the-practice, we group requirements and solutions along four distinct axes of classification:
where the system is deployed, which quantum-hardware
modality is being integrated, how applications use it, and which
quantum-computing era, \ac{NISQ} or \ac{FTQC} (or both), it must support.

\subsection{Axis 1: Deployment Environment}
The two primary \textbf{deployment environments} for QHPC stacks are cloud and
on-premises operation. Each imposes different integration, orchestration,
latency, and ownership constraints, and a stack must support both without
architectural rework as well as compositions of the two across administrative
trust boundaries (often referred to as federated operation).

In \textbf{on-premises} deployments, the \ac{HPC} center is responsible for integrating
the quantum system into the local facility environment, including resource
management, security policy, data movement, and user access paths~\cite{Mansfield_Int,qrmi-slurm,iqm_eqxa_2025}. This creates
practical design questions beyond merely attaching a device. One issue is how
the quantum resource is presented to users through the center's software stack.
Another is the placement of classical compute beside the
quantum resource to handle compilation, runtime services, orchestration, \ac{QEC}
and control. The center must then decide whether those nearby classical
resources are dedicated to quantum operation or shared with other workloads.

In the \textbf{cloud} model, quantum resources may be consumed from an
\ac{HPC}-center perspective as a form of capability extension.
From the outside, the resource may appear to be an external endpoint. On the provider side, however, the environment still resembles a datacenter
with its own orchestration stack, service boundaries, and classical execution tiers.
In practice, these environments often rely on service-managed and
containerized execution models rather than traditional \ac{HPC} scheduler-native integration;
for example,
Amazon Braket Hybrid Jobs supports provider-managed classical instances and custom container images~\cite{braket-hybrid-jobs,braket-reservations}.

\textbf{Federated operation} is best understood as a composition of the two
primary models across multiple administrative trust boundaries rather than as
a third independent category. The functions it depends on: authentication,
authorization, scheduling, and data movement, are required in all
deployments, but federated settings demand that they interoperate across
sites under coordinated policy. Much of that coordination lies outside the
stack's scope and is the responsibility of the participating operators; the
stack's responsibility is to provide the hooks (pluggable identity providers,
resource description portable across schedulers, secure data paths) that
make such cross-domain operation possible.

\subsection{Axis 2: Quantum Hardware Modality}
Quantum computers can be built based on a wide variety of different implementations of quantum mechanical principles. This is usually referred to as its modality. Among the wide range of approaches that exist, Diamond Vacancies, Neutral/Cold Atoms, Photonic Systems, Spin Qubits, Superconducting Qubits, and Trapped Ions have to date received the most attention. These modalities impose different latency, control, and orchestration requirements, especially when considering \ac{QEC}, whose operations must complete within the coherence time of the qubits.
As a consequence, current software stacks typically target a single modality, or a small subset, to simplify optimization and control.
However, long-term scalability will require well-defined abstraction layers that decouple higher-level software from hardware-specific details. A flexible tool and compiler pipeline capable of adapting transformations, lowering, and optimization to modality-specific constraints while preserving a consistent interface to upper layers is also required.

\subsection{Axis 3: Application Interaction Patterns}
The type of quantum workload shapes the requirements a \ac{QHPC} software
architecture must adhere to. For that, this paper identifies two meta-patterns, workflow-centric and accelerator-centric.
Both are relevant across cloud and on-premises deployments.

\subsubsection{Workflow-Centric Pattern}

In the workflow-centric pattern, applications consist of independent classical and quantum stages, each maintaining its own state, if any. They are separated by workflow boundaries and
coordinated by a scheduler or workflow engine. Direct interaction between
these stages from within the execution of a stage is explicitly
excluded, as data/control exchange occurs only at stage boundaries rather
than through tight runtime coupling. Typical workflow systems treat HPC and QPU resources equally, delegating work to whichever system is best suited. There is no requirement or implication that multiple workflow systems are used or that HPC needs to have a workflow system separate from the one coordinating quantum and classical hybrid workflow.

This pattern is well-suited for execution across sites, allowing quantum steps to be offloaded to remote resources, including cloud-based systems, without requiring low-latency coupling to the \ac{HPC} stages. It also applies to on-prem configurations where the quantum stage remains a distinct part of a larger workflow.

\subsubsection{Accelerator-Centric Pattern}

In the accelerator-centric pattern, applications are driven by a larger \ac{HPC} program that invokes quantum kernels during execution, analogous to GPU accelerator offload. These kernels are typically fine-grained and dispatched on demand, with \ac{HPC} and quantum resources co-allocated to enable low-latency interaction and efficient execution. This requires careful consideration to avoid HPC idling or timeout behavior

Unlike the workflow-centric model, this pattern requires tighter coupling between \ac{HPC} and \ac{QPU}, support for shared state across kernel invocations, and frequent synchronization or runtime feedback, including mid-circuit interaction when supported. Efficient utilization also motivates multi-tenant resource sharing through a second-level quantum scheduler, enabling independent kernel scheduling with explicit \ac{QoS} and bounded-progress guarantees.

\subsection{Axis 4: Targeted Quantum Computing Era}
While current systems target \ac{NISQ} workloads, future systems will support \ac{FTQC}, with both eras expected to coexist for some time. Software stacks, therefore, differ in which era they support and how much of this distinction is exposed to user-facing layers. In the \ac{NISQ} era, application execution often includes algorithm-specific hybrid loops (e.g., \ac{VQE}) that are performance-sensitive but not strictly timing-critical. In \ac{FTQC}, execution introduces inner \ac{QEC} loops where syndrome readout and correction must operate under strict timing constraints and affect the correctness of execution.

This shift can be understood through explicit timing-domain separation. The \ac{PTD}~\cite{nvqlink-arxiv} captures waveform-level interactions with device physics, the \ac{DTD} handles hardware-local control and fast feedback, and the \ac{RTD} supports bounded-latency classical processing, such as decoding and adaptive control. Higher software layers operate in the \ac{ATD}, relying on abstractions that isolate these timing constraints.

This separation motivates a runtime interface that abstracts application execution from hardware-specific details, allowing the same application model to span \ac{NISQ} and \ac{FTQC} systems. At the same time, lower-level hardware capabilities, such as conditional control, parameter streaming, and feedback primitives, should be exposed through stable, queryable APIs, enabling higher layers to make informed execution and compilation decisions without embedding device-specific logic. This balance allows the architecture to remain stable while accommodating evolving \ac{FTQC} requirements.

\section{First Established Cross-Stack Interfaces}

Before turning to complete software stacks, we highlight two emerging interface community efforts that address recurring integration problems across the \ac{QHPC} landscape: \ac{QDMI} and \ac{QRMI}.

Their emergence and uptake reflect a broader trend towards decoupled and interoperable interfaces.

\subsection{Quantum Device Management Interface (QDMI)}
\ac{QDMI}~\cite{wille2024qdmi} defines a C-level standard for interacting with quantum devices, providing a backend-agnostic interface for device discovery, capability queries, job lifecycle management, and result retrieval. It provides a device-facing abstraction boundary in QHPC systems, allowing upper layers to interact with quantum hardware without embedding vendor-specific APIs. In this role, QDMI exposes structured hardware information that compiler toolchains and runtime systems can use, while remaining independent of higher-level resource management policies.

\ac{QDMI} is built around a layered lifecycle model supporting authenticated, multi-tenant device sessions. It exposes hardware properties at device, qubit, and operation levels, enabling hardware-aware compilation and informing scheduling decisions without requiring hardware-specific logic in upper layers. It also supports asynchronous execution tracking and low-latency interaction, allowing integration into both workflow-centric and accelerator-centric execution patterns while remaining focused on device interaction rather than resource allocation semantics.

\subsection{Quantum Resource Management Interface (QRMI)}

\ac{QRMI} \cite{qrmi-slurm} is a Rust-based, vendor-neutral interface that provides Rust, C, and Python APIs for interacting with quantum resources. It abstracts resource acquisition, managed execution, resource selection, allocation, job submission, and lifecycle management—creating a resource-management boundary in QHPC systems. This allows upper layers to interact with quantum resources as schedulable entities without embedding site-specific or scheduler-specific control logic. By doing so, \ac{QRMI} enables applications, schedulers, and resource managers to coordinate execution while remaining independent of specific device abstractions and compiler toolchains.

\ac{QRMI} integrates with existing resource-management systems through dedicated plugin libraries, supporting Slurm, PBS, Open Cluster Scheduler, Flux, and IBM Spectrum Load Sharing Facility (LSF). By describing quantum resources using the same generic framework as other accelerators like GPUs, \ac{QRMI} enables standard scheduling, allocation, cost management, and policy enforcement. This unified approach supports both workflow-centric and accelerator-centric execution patterns while maintaining consistent lifecycle semantics, and works seamlessly across cloud and on-premises QPU deployments. Rather than focusing on structured device abstraction, \ac{QRMI} prioritizes resource management and execution. This design philosophy enables multiple front-ends (Qiskit, Pulser, Cuda-Q) to leverage \ac{QRMI} without requiring framework-specific implementations. Additionally, QRMI enables job submission to simulators either through cloud connections or by piping hardware specifics to local simulators, demonstrating its adaptability across different execution environments without impacting workflow design.

\section{Classifying and Mapping Existing \ac{QHPC} Software Stacks}
\label{sec:stacks}
In this rapidly growing area of \ac{QHPC}, a wide variety of software stacks have been developed independently by vendors and academic researchers alike.
Prior surveys have examined this landscape from a literature perspective: D\"obler and Jattana~\cite{dobler2025survey} categorize 107 publications spanning hybrid hardware architectures and software stacks, while Elsharkawy et al.~\cite{elsharkawy2025integration} review quantum programming tools and their suitability for \ac{HPC} integration. Our work is complementary: rather than surveying the literature, we examine production \ac{QHPC} stacks as deployed today, and from that state-of-the-practice derive the common design patterns and interface boundaries that motivate a shared reference architecture.
In order to understand their impact on the
current state-of-the-practice and to map a path to a more unified ecosystem, we explore and classify nine of the most common and available stacks, and derive commonalities and differences. For each stack we describe the architecture, its source and use cases, and classify them along the four axes introduced in \autoref{sec:tec-req}. While this survey does not encompass the entire industry, it is sufficiently comprehensive to identify common design patterns which guide a shared architecture.

\begin{figure}[h]
    \centering
    \includegraphics[width=0.6\textwidth]{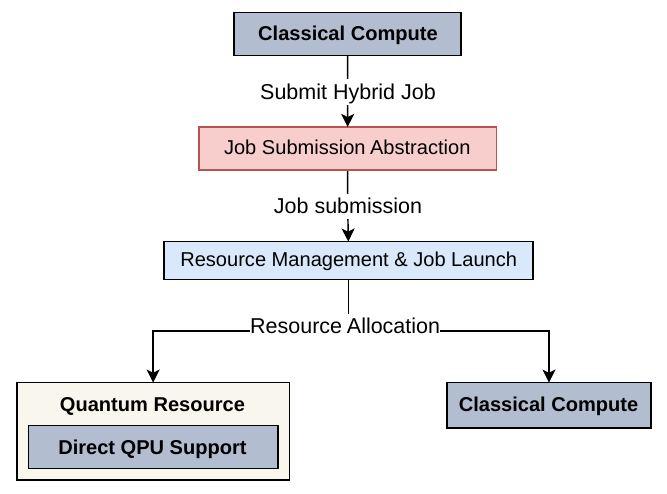}
    \caption{Common hybrid job submission workflow across all surveyed
    stacks. A Job Submission Abstraction decouples classical compute
    from the resource manager, which allocates quantum and classical
    resources in parallel. The quantum resource encapsulates its own
    classical QPU support layer, isolating hardware-specific
    control from upper stack layers.}
    \label{fig:openqse-workflow}
\end{figure}

To further support the comparison, we map the workflow of each stack to a
common hybrid job
submission pattern, which we observed across all surveyed stacks and which is illustrated
in \autoref{fig:openqse-workflow}: a classical
compute node submits a hybrid job through a Job Submission Abstraction
layer, which decouples the application from the underlying resource
manager. The Resource Management and Job Launch layer then allocates
both quantum and classical resources concurrently or separately, depending on the user requirements, with the quantum
resource encapsulating its own classical \ac{QPU} support layer.

In the following, we describe each of the nine stacks in our study, listed in alphabetical order and labeled based on the name of the respective developing company or project.

\subsection{Stack 1: Amazon~Web~Services}
\label{stack-aws}

\ac{AWS} supports several \ac{QHPC} deployment models: cloud, on-premises, federated, and mixed configurations. Cloud deployments range from single-service solutions using Amazon Braket, the AWS quantum computing service, to multi-service architectures combining Braket with \ac{HPC} services. On-premises and federated deployments can integrate with Braket or leverage cloud \ac{HPC} for on-premises quantum systems~\cite{zhao2025}. 

Amazon Braket provides unified access to \ac{QPU} technologies from multiple hardware providers, including AQT, IonQ, IQM, QuEra, and Rigetti. The service accepts OpenQASM~3~\cite{cross2022openqasm} circuit representations (including the OpenPulse dialect for pulse-level access) and Analog Hamiltonian Simulation programs. The Amazon Braket SDK has open-source plugins for CUDA-Q, PennyLane, and Qiskit integration. Braket exposes device capability metadata including calibration and fidelity data when available from \ac{QPU} vendors, informing optional compiler optimization. Precompiled circuits can execute without optimization, and parametric circuits are cached for faster batch execution. Quantum circuits can be scheduled via on-demand access, priority access through Amazon Braket Hybrid Jobs, or exclusive access with reservations through Braket Direct. Fair-share queuing and \ac{HPC} queue integration (e.g., \ac{SPANK} plugins) have not yet been implemented. Execution queues are resiliently maintained, and results persist in customer-owned Amazon S3 buckets. 

Amazon Braket Hybrid Jobs supports single-service cloud \ac{QHPC} by providing containerized execution environments that manage classical infrastructure lifecycle and quantum device integration. Multi-service and mixed deployments can leverage purpose-built \ac{HPC} environments on Amazon EC2 clusters provisioned from more than 850 instance types~\cite{amazon-ec2} optimized to optimized for different computing requirements. Cluster infrastructure can be service-managed~\cite{aws-pcs} or self-managed~\cite{aws-pcluster}. Cloud-native workflow orchestration~\cite{aws-stepfunctions} and batch scheduling~\cite{aws-batch} services provide alternatives with reduced operational overhead. These approaches support both workflow-centric and accelerator-centric patterns. Multi-service and mixed deployments allow the pairing of external \ac{QHPC} software stacks with the Braket circuit execution model. Recently, the Amazon Braket implementation of the \ac{QDMI} specification~\cite{wille2024qdmi} enabled the integration of Braket into every \ac{QDMI}-compatible \ac{QHPC} software stack. 

\ac{AWS} supports federated cross-site access with external identity providers, fine-grained authorization with role-based and attribute-based access control, and protection of data at rest and in transit including VPN and \ac{AWS} Direct Connect for private network connections.

\subsection{Stack 2: IBM}
\label{stack-ibm}

IBM Quantum systems are based on superconducting qubit technology, which enables fast gate operations while imposing stringent requirements on timing, calibration, and control. Access to these systems is provided through multiple deployment models offered under the \emph{IBM Quantum Platform}, including cloud-based access via the \emph{Quantum Cloud API} and on-premises deployments accessed through the \emph{Quantum System API}.

IBM uses the \ac{QRMI}~\cite{qrmi-slurm}, which supports both APIs, and integrates \ac{QRMI} with the Slurm workload manager via its \ac{SPANK} plugin.

At the programming level, Qiskit serves as IBM’s quantum SDK. Qiskit provides a Python interface as well as a C-API, which enables the use of additional programming environments, including C++ and Julia. In HPC‑integrated deployments, Qiskit‑based applications execute on classical HPC systems and access quantum systems via IBM's \ac{QRMI}, allowing quantum execution to be incorporated into HPC workflows while keeping application logic decoupled from system‑level resource management. Using this architecture, IBM has demonstrated both workflow-centric and accelerator-centric application patterns \cite{closed-loop}\cite{ibm-workflow}. 

In the workflow-centric pattern, classical HPC jobs and quantum jobs are orchestrated by a workflow engine and executed as distinct stages within a larger pipeline, with loose temporal coupling between classical and quantum execution \cite{ibm-workflow}. In contrast, the accelerator-centric pattern enables tighter integration between HPC and quantum systems, where classical and quantum executions are more closely coordinated and can partially overlap, improving overall system utilization and enabling closed-loop hybrid execution \cite{closed-loop}.

Taken together, these deployments correspond to the ``quantum as a co-processor to HPC'' model described in IBM’s \ac{QCSC} reference architecture~\cite{ibm-reference-arch}. While this model represents the dominant integration approach in current deployments, the \ac{QCSC} reference architecture also outlines future phases with more tightly integrated execution models, including near-time coupling and support for outer decoders.

\subsection{Stack 3: IonQ}
\label{stack-ionq}

IonQ provides access to trapped-ion \ac{QPU} resources through its
\emph{Quantum Platform}~\cite{ionq-quantum-cloud, openqse_wg_ionq_2026}, which spans scheduling, compilation, tenant
management, and observability. It integrates with
Qiskit~\cite{qiskit}, CUDA-Q~\cite{cudaq2026}, and
PennyLane~\cite{pennylane2022} SDKs. The primary deployment model is
cloud-managed; on-premises and federated access pathways are emerging
via \ac{QDMI}~\cite{wille2024qdmi} and \ac{QRMI}~\cite{qrmi-slurm}
plugin implementations~\cite{openqse-workshops}.

Scheduling is multi-tiered~\cite{openqse-workshops}: a platform tier
manages tenant allocations via proportional fair-share; a \ac{QPU} tier
interleaves user jobs with calibrations, which consume accountable \ac{QPU}
time; and a hardware tier sequences low-level instructions. The control
electronics layer surfaces timing information to upper tiers, and
\autoref{sec:control-electronics} discusses representative IonQ Forte
gate-execution timings~\cite{ionq-forte} and their cross-stack
architectural implications.

Compilation applies device-agnostic optimization then device-specific
lowering into the native gate set, guided by calibration
data~\cite{openqse-workshops}. \ac{QRMI}~\cite{qrmi-slurm} exposes
IonQ as a schedulable backend within HPC workload managers, while
\ac{QDMI}~\cite{wille2024qdmi} provides the device-facing interface
that the runtime layer uses to interact with IonQ hardware. This
layering supports consistent operation against cloud-hosted IonQ
deployments and could extend to on-premises IonQ deployments.

IonQ supports the workflow-centric pattern via direct API access,
cloud marketplaces~\cite{ionq-quantum-cloud}, and large-scale
campaigns such as QC-AFQMC~\cite{zhao2025} coordinating IonQ
\acp{QPU} with NVIDIA GPUs via Amazon Braket. \ac{FTQC}
support is on IonQ's roadmap, with the multi-tier control architecture
providing the timing primitives needed for syndrome dispatch and
measurement feedback in the deterministic and real-time domains.

\subsection{Stack 4: JHPC-Quantum: Supercomputer Quantum Computer Computation (SQC) scheduler and ecosystem}
\label{stack-riken}

The JHPC-Quantum Project~\cite{openqse_wg_jhpc_2026}, jointly developed by RIKEN, Softbank, the University of Tokyo, and University of Osaka targets an HPC-oriented integration of quantum and classical computing.
Its primary objective is to enable efficient utilization of both HPC and QC resources, with particular emphasis on ensuring that large-scale HPC resources are not left idle or waiting on quantum processing. At the same time, standalone QC jobs are executed with lower priority, leveraging otherwise idle quantum resources to maximize overall system utilization. To this end, a lightweight scheduler called SQC scheduler is introduced between HPC and QC, designed to prioritize and promptly return results for requests originating from HPC workloads.

This scheduling mechanism not only orchestrates the flow of requests to quantum devices, but also abstracts away heterogeneity among backend quantum systems. Since users do not directly access individual QC backend servers, differences in their interfaces are effectively hidden. As a result, users are required to interact only with a unified scheduler interface, significantly simplifying the development and execution of hybrid quantum-HPC workflows.

By extending the execution layers of existing SDKs, such as Qiskit, TKET, and CUDA-Q to interface with the proposed scheduler, quantum programs written in these frameworks can also be executed without modification.

\subsection{Stack 5: Munich Quantum Software Stack (MQSS)}
\label{stack-mqss}

The MQSS, developed as part of the Munich Quantum Valley (MQV), is a multi-layered architecture designed to integrate heterogeneous quantum hardware into \ac{HPC} environments~\cite{mqss}. It was started specifically as an open source solution with multi-modality support to match the developments in the MQV, but it also aims at broad applicability beyond MQV. As such, it supports both cloud-based and on-prem deployments, including tightly coupled integration within HPC systems and more loosely coupled remote-access configurations~\cite{openqse_wg_mqss_2026}.

The architecture is organized into three layers.
The front-end layer provides application-facing interfaces and supports frameworks such as Qiskit and PennyLane, domain-specific languages, as well as specific hybrid programming approaches, like QPI~\cite{qpi-mqss} or OpenMP integration~\cite{openmp-mqss}.
The middle-end layer integrates quantum workflows into the \ac{HPC} environment and is responsible for compilation, resource allocation, and orchestration of hybrid applications.
It supports both workflow-centric and accelerator-centric execution patterns.
This layer is supported by tooling from the \ac{MQT}~\cite{mqt}, including benchmarking (MQT Bench~\cite{quetschlich2023mqtbench}), compilation across qubit modalities (MQT QMAP~\cite{wille2023qmap}), \ac{MLIR}-based compilation infrastructure (MQT Core~\cite{burgholzer2025MQTCore}), and scheduling and resource selection (MQT Predictor~\cite{quetschlich2025mqtpredictor}).
Integration with \ac{Slurm} is provided, including support for multi-resource scheduling through a second-level scheduler. The back-end layer interfaces with quantum hardware through open interfaces, particularly \ac{QDMI} initially developed by the MQSS team~\cite{wille2024qdmi}, enabling interaction with multiple \ac{QPU} backends.

MQSS has been deployed in an \ac{HPC} setting at the Leibniz Supercomputing Centre, where it is used to manage a tightly coupled integration with a superconducting quantum device~\cite{Mansfield_Int}.
The stack exposes telemetry from both quantum resources and the surrounding \ac{HPC} environment.
It is currently used in \ac{NISQ}-focused deployments, but includes architectural support for future \ac{FTQC} operation, such as consideration of control electronics integration within the resource-management layer.

\subsection{Stack 6: Pasqal}

Pasqal develops and maintains a \ac{QHPC} integration approach for its neutral-atom systems~\cite{henriet2020} that combines on-premises and cloud deployment support with explicit integration into HPC resource managers and a second device-level scheduling layer~\cite{pasqal-hpc,openqse_wg_pasqal_2026,pasqal_cloud_hybrid_algos}.

Pasqal relies on its \ac{QRMI}~\cite{qrmi-slurm} implementation to integrate with the HPC system scheduler and reuses it to provide framework-independent runtime access to its quantum resources,
supporting its native Pulser~\cite{pulser}, Qiskit primitives, and CUDA-Q~\cite{cudaq2026} as programming frontends across on-premises integration, cloud bursting, and other mixed deployment modes.
Pasqal supports both accelerator- and workflow-centric execution models.

Below the HPC resource manager, a second-level quantum-aware scheduler manages QPU-specific constraints that are difficult to capture directly in standard HPC schedulers, such as calibration-aware execution, and enforces center-specific usage policies. While the QPU itself must be allocated to a single active user or job at a time, this layer enables multiple user sessions at the access layer. It performs access control, manages execution windows, and can improve device utilization through mechanisms such as time-slicing and shot limits, while keeping the QPU's internal queue aligned with HPC scheduler policies such as priorities and maximum walltime.

Pasqal also exposes telemetry and observability capabilities through administrative interfaces in its on-premises stack, supporting system monitoring and operational visibility as described in~\cite{pasqal-hpc}.

\subsection{Stack 7: Quantinuum}
\label{stack-quantinuum}

Quantinuum trapped-ion quantum computers are available via cloud access or via on-premise deployments. The QPUs are self-contained and capable of using any QEC code. The software stack includes tools for application development, QPU programming, and compilation. Cloud hosted quantum computers are accessed via the Nexus~\cite{quantinuum_nexus} API while on-premise systems have an option for direct gRPC-based~\cite{grpc} access.

The programming language Guppy~\cite{Koch2024, Koch2025} allows programmers to seamlessly combine classical and quantum logic in real-time  with full support for measurement-based program flow control. Other development frameworks including Q\#~\cite{Svore_2018} and CUDA-Q~\cite{cudaq2026} are supported via the industry standard \ac{QIR}~\cite{QIRSpec2021} format. TKET optimizing compiler performs quantum optimizations on the intermediate representation and QPU targeting.

Quantinuum QPU scheduling is based on an adaptive time-sliced fair queuing approach. User job slices are scheduled just-in-time based on priority, order of submission, and previous execution history within a fairness window, alongside other scheduling attributes. Automated calibration routines are inserted between user job slices by the scheduler when necessary.

In HPC deployments, Quantinuum focuses on the workflow-centric pattern emphasizing complex interactions between HPC and QPU resources deployed side by side. Quantinuum workflow engine Tierkreis~\cite{tierkreis} is a task-based, asynchronous orchestrator for heterogeneous hardware types. It is programming language agnostic, allowing reuse of existing HPC software and tools. 

The fundamental design principle is a worker-executor model. Workers provide atomic tasks to a workflow. Executors run assigned tasks on a target hardware, be it cloud, HPC or QPU. Currently, Tierkreis supports executors for \ac{Slurm}, \ac{PBS} and pjsub on the HPC side; and Nexus and direct QPU submission on the quantum side. An example of a hybrid biochemistry application using Tierkreis combines existing HPC chemistry application with quantum kernels in a complex workflow~\cite{yamamoto2026}.

\subsection{Stack 8: Quantum Brilliance}
\label{stack-qb}

Quantum Brilliance (QB) provides \ac{QHPC} capabilities through its \ac{QTIL}~\cite{openqse_wg_qb_2026}. Implemented in Go (Golang), \ac{QTIL} is responsible for quantum resource allocation and task scheduling, interfacing with the \ac{QPU} web server via API calls. 

Within the QB framework, \ac{QTIL} integrates with the client SDK and the \ac{QDK}, where user-defined quantum circuits or pulse-level instructions are submitted via API. The \ac{QDK} comprises the \ac{QPU} and QCStack, a software stack that includes a web server and supporting libraries for functionalities such as API handling, transpilation, optimization, and calibration. In this context, \ac{QTIL} operates as the resource management layer, orchestrating access to \acp{QPU} exposed through the QCStack interfaces.

Importantly, \ac{QTIL} is not tightly coupled to the QB software stack. Its design allows it to interface with alternative client SDKs and backend systems, provided that compatible API endpoints are available. This makes \ac{QTIL} adaptable to different software environments and hardware platforms.

\ac{QTIL} manages resource allocation on a per-\ac{QPU} basis, allowing users to specify parameters such as allocation start time, duration, and the number of \acp{QPU} required. \ac{QPU} metadata is maintained in a relational SQL database (e.g., MySQL), enabling structured classification of resources based on configurable attributes. These attributes can be used to define logical groupings of \acp{QPU} and to enforce allocation policies.

Resource allocation is secured using \ac{JWT}s in combination with secret keys stored in the database.
For integration into \ac{HPC} environments, \ac{QTIL} provides support for \ac{Slurm}~\cite{slurm-scontrol}, enabling hybrid quantum-classical workflows. In contrast to approaches based on the \ac{SPANK} plugin, \ac{QTIL} adopts a license-based model using the Lua plugin. In this setup, users request classical resources (CPU/GPU) through \ac{Slurm} as usual, while additionally specifying \ac{QPU} requirements. This design minimizes changes to existing \ac{Slurm} configurations, as modifications are primarily limited to prolog and epilog scripts.

\subsection{Stack 9: Xanadu}
\label{stack-xanadu}

The Xanadu quantum stack~\cite{openqse_wg_xanadu_2026} features PennyLane~\cite{pennylane2022}, a platform-agnostic frontend, and Catalyst~\cite{Ittah2024}, its dedicated compiler~\cite{openqse-workshops}. PennyLane maps high-level applications to structured hybrid programs, which Catalyst progressively lowers to device-specific operations for execution via a runtime layer. Catalyst compiles these workloads using an \ac{MLIR}~\cite{mlir} toolchain, with  classical and quantum instructions combined in a single optimized program. Through tools, like the Jeff framework~\cite{Jeff}, and a comprehensive plugin interface, the ecosystem natively interoperates with Qiskit~\cite{qiskit}, Cirq~\cite{cirq}, Amazon Braket~\cite{braket-hybrid-jobs}, and other vendor toolchains.

Xanadu has developed demonstrators of its quantum system technologies, including X8~\cite{arrazola2021quantum}, Borealis~\cite{Madsen2022Nature}, and Aurora~\cite{Bourassa2025Scaling}, while current development is focused on a utility-scale quantum computing system. Future access is expected to center on cloud and federated models delivered through remote PennyLane environments. Software deployment is highly flexible, with support for pre-built Python packages, customized Docker containers, Spack integration~\cite{gamblin2015spack}, and local source builds.

As Catalyst enables targeted heterogeneous compilation from a single user workload, both accelerator-centric and workflow-centric patterns are supported. A modular C++ runtime uses stubs for quantum instructions, tailoring execution to any target platform. This unified classical-quantum model natively supports local execution, remote submission, and directly integrated heterogeneous environments. Classical code components (control flow, memory semantics, linear algebra) compile directly for CPUs and accelerators (GPUs, TPUs) using \ac{MLIR}/LLVM~\cite{lattner2004llvm} backends and ecosystem libraries like JAX~\cite{jax2018github}, ROCm\texttrademark~\cite{hip-docs}, and CUDA~\cite{nvidia_cuda_platform}.

While the software delegates workload-level scheduling and device provisioning to the surrounding system context, Catalyst provides the necessary tie-ins for these execution models. Furthermore, as Xanadu prepares for \ac{FTQC} designs, Catalyst is actively integrating logical and physical layer quantum error correction passes into its lowering stack~\cite{catalyst-dialects}.

\section{Design Patterns \& Future Features}

Across the current stack landscape, several design patterns emerge. The most
consistent is a layered separation between applications, runtime services,
compilation, and hardware-facing control. This structure allows adaptation
across deployment environments without replacing the full stack, while enabling
vendor specialization near the device behind more stable upper interfaces.
Another pattern is the use of explicit runtime boundaries for submission,
status tracking, and result return. Although not yet portable across stacks,
these interfaces already simplify integration with workflows and facility
software.

A staged view of program preparation is also common. Quantum programs are
transformed and lowered through multiple compilation steps prior to execution,
providing a natural insertion point for backend-specific optimization without
exposing hardware details to applications. Reservation handling, scheduler
integration, and capability discovery are increasingly treated as standard
operational features. These capabilities enable coordination of shared
resources, visibility into system state, and more predictable execution in
hybrid environments.

Telemetry is similarly established across stacks. Systems routinely collect
run-state information, operational health signals, execution metadata, and
hardware-facing measurements for debugging, scheduling, and operations. While
the specific data varies across modalities and vendors, its consistent presence
indicates that observability is an architectural requirement rather than an
optional feature.

The next step is to consolidate these patterns into common interfaces. An
application-facing runtime API would provide a stable offload contract and
reduce dependence on stack-specific SDK entry points. A composable tool and
pipeline model would allow compilation and transformation passes to be reused
across frameworks instead of remaining isolated. Connection management requires
similar abstraction, enabling transport and fabric differences to be handled
without impacting upper layers. Communication semantics must also extend beyond
one-way submission and result retrieval to support bidirectional interaction,
allowing quantum-side services to trigger adaptation, event notifications, and
resource requests.

Shared usage is already a common operating mode, driven by cloud providers that
must support concurrent access from many users and therefore incorporate
multi-tenant scheduling into their stacks. However, these mechanisms remain
fragmented, and a unified approach to policy-aware sharing is needed as similar
patterns extend into HPC environments. Calibration visibility is closely
related, as maintenance and retuning directly affect scheduling decisions.
Rather than enforcing a fixed telemetry model, a more durable goal is a standard
mechanism for accessing system information that can evolve in a backward-compatible
way as new hardware signals and control features emerge.

Finally, hardware capability exposure will become increasingly important as
systems progress toward stronger error mitigation and fault-tolerant operation.
Control-electronics features, hardware-near classical processing, and feedback
paths should be surfaced through structured interfaces, so higher layers can
make informed compilation, reservation, and execution decisions.

\providecommand{\fpresent}{\checkmark}
\providecommand{\fpartial}{\checkmark}
\providecommand{\fnone}{}

\section{Discussion and Mapping to Common Blueprint}

As shown in the previous section, the different stacks target different scenarios and support a wide range of properties and systems. Mapped to the classification axes, they do provide wide coverage across the entire design space. At the same time and despite their differences, they roughly map to our common layer model in \autoref{fig:openqse-architecture}.

\autoref{tab:stack-convergence} reorients the mapping around \emph{interfaces and components}: each row is an interface or software component, each column a stack, with cells indicating support (\fpresent), either directly or via a translation layer, or absence of support (\fnone). Solid rows in \autoref{tab:stack-convergence} mark where the community is heading towards convergence (e.g., OpenQASM~3); sparse rows mark the boundaries that are still consolidating.

\begin{table*}[t]
\centering
\footnotesize
\setlength{\tabcolsep}{4pt}
\renewcommand{\arraystretch}{1.05}

\begin{tabular}{
|>{\raggedright\arraybackslash}p{0.24\textwidth}
|*{9}{>{\centering\arraybackslash}p{0.055\textwidth}|}
}
\arrayrulecolor{black}
\hline
\rowcolor{gray!30}
\textbf{Interface / standard} &
\textbf{AWS} & \textbf{IBM} & \textbf{IonQ} & \textbf{JHPC-Q} &
\textbf{MQSS} & \textbf{Pasqal} & \textbf{QB} & \textbf{Qntm} & \textbf{Xan} \\
\hline

\multicolumn{10}{|l|}{\cellcolor{gray!12}\textit{Deployment Model}} \\
\hline
Cloud
& \fpresent & \fpresent & \fpresent & \fpresent & \fpresent & \fpresent & \fnone & \fpresent & \fpresent \\ \hline
On-premises
& \fnone & \fpresent & \fnone & \fpresent & \fpresent & \fpresent & \fpresent & \fpresent & \fnone \\ \hline

\multicolumn{10}{|l|}{\cellcolor{gray!12}\textit{Resource-Level Scheduling}} \\
\hline
\acs{Slurm} (\acs{SPANK} / GRES / Lua)
& \fpresent & \fpresent & \fnone    & \fpresent    & \fpresent & \fpresent & \fpresent & \fpresent & \fnone    \\ \hline

\multicolumn{10}{|l|}{\cellcolor{gray!12}\textit{SDK Interface}} \\
\hline
Qiskit~\cite{qiskit}
& \fpresent & \fpresent & \fpresent & \fpresent & \fpresent & \fpresent & \fpresent    & \fpresent & \fpresent \\ \hline
PennyLane~\cite{pennylane2022}
& \fpresent & \fpresent    & \fpresent & \fpresent    & \fpresent & \fnone    & \fpresent    & \fpresent    & \fpresent \\ \hline
TKET
& \fnone    & \fpresent    & \fpresent    & \fpresent & \fpresent    & \fnone    & \fpresent    & \fpresent & \fpresent    \\ \hline
CUDA-Q~\cite{cudaq2026}
& \fpresent & \fpresent    & \fpresent & \fpresent & \fpresent    & \fpresent & \fpresent    & \fpresent & \fpresent \\ \hline

\multicolumn{10}{|l|}{\cellcolor{gray!12}\textit{Quantum-Task Compiler / IR}} \\
\hline
\acs{MLIR}-based lowering~\cite{mlir}
& \fnone    & \fnone    & \fnone    & \fnone    & \fpresent & \fnone    & \fnone    & \fnone    & \fpresent \\ \hline
Vendor-specific compiler
& \fpresent    & \fpresent    & \fpresent    & \fnone    & \fnone & \fpresent    & \fpresent    & \fpresent    & \fpresent \\ \hline
OpenQASM~\cite{cross2022openqasm}
& \fpresent & \fpresent & \fpresent & \fpresent & \fpresent & \fpresent & \fpresent & \fpresent & \fpresent \\ \hline
\acs{QIR}~\cite{QIRSpec2021}
& \fnone    & \fnone    & \fpresent  & \fnone    & \fpresent    & \fnone    & \fnone    & \fpresent & \fnone    \\ \hline

\multicolumn{10}{|l|}{\cellcolor{gray!12}\textit{Resource / Device Interface}} \\
\hline
\acs{QDMI}~\cite{wille2024qdmi} (device, MQT)
& \fpresent & \fnone    & \fpresent & \fnone    & \fpresent & \fnone    & \fnone    & \fnone    & \fnone    \\ \hline
\acs{QRMI}~\cite{qrmi-slurm} (resource)
& \fnone    & \fpresent & \fpresent & \fnone    & \fnone    & \fpresent & \fnone    & \fnone    & \fnone    \\ \hline

\multicolumn{10}{|l|}{\cellcolor{gray!12}\textit{Quantum-Task Level Scheduling}} \\
\hline
*Stack-specific scheduling
& \fpresent & \fpresent    & \fpresent & \fpresent & \fpresent    & \fpresent & \fpresent & \fpresent & \fpresent \\ \hline

\end{tabular}

\vspace{1mm}
\caption{Interface convergence across surveyed stacks (April 2026).
\fpresent{} indicates support (direct or via translation); blank
indicates absence. All stacks expose vendor-defined REST/gRPC backends
and support federated operation via external identity providers (rows
omitted). *Stack-specific scheduling: a second-level scheduler beyond
the per-device queue.}
\label{tab:stack-convergence}
\end{table*}

\section{Introducing the openQSE Architecture}

Our survey has shown the wide variety of current state-of-the-practice software solutions for various deployment scenarios, hardware modalities, interaction patterns, and targeted quantum era, as well as the different states in their development. At the same time it is also clear that not all stacks cover the entire scenario and that cross-stack support is sought after in many solutions. In particular, the establishment of the two interface approaches, \ac{QDMI} and \ac{QRMI}, shows a need to converge towards compatible approaches to further drive the ecosystem.

To accomplish this, we need a reference architecture, which allows us to map each stack to it, identify matching components, and extract interfaces that ultimately could lead us to community standards supporting interoperable stacks. Following the results of our survey, we propose
\ac{openQSE},
a community-driven architecture and specifications
effort~\cite{openqse-workshops} to establish such a reference architecture. The goal is explicitly not
to replace existing SDKs, vendor
stacks, or open source efforts.  Instead, the goal of this initiative is to define
vendor-neutral interface boundaries so applications, runtimes, resource
managers, and backend implementations can interoperate while still
allowing implementation-specific or site-specific optimizations. This will enable the community to connect the different software stacks, exploit cross-stack functionality, and ultimately make all functionality available across the entire \ac{QHPC} ecosystem.

\begin{figure*}[t]
      \centering
      \includegraphics[width=\textwidth]{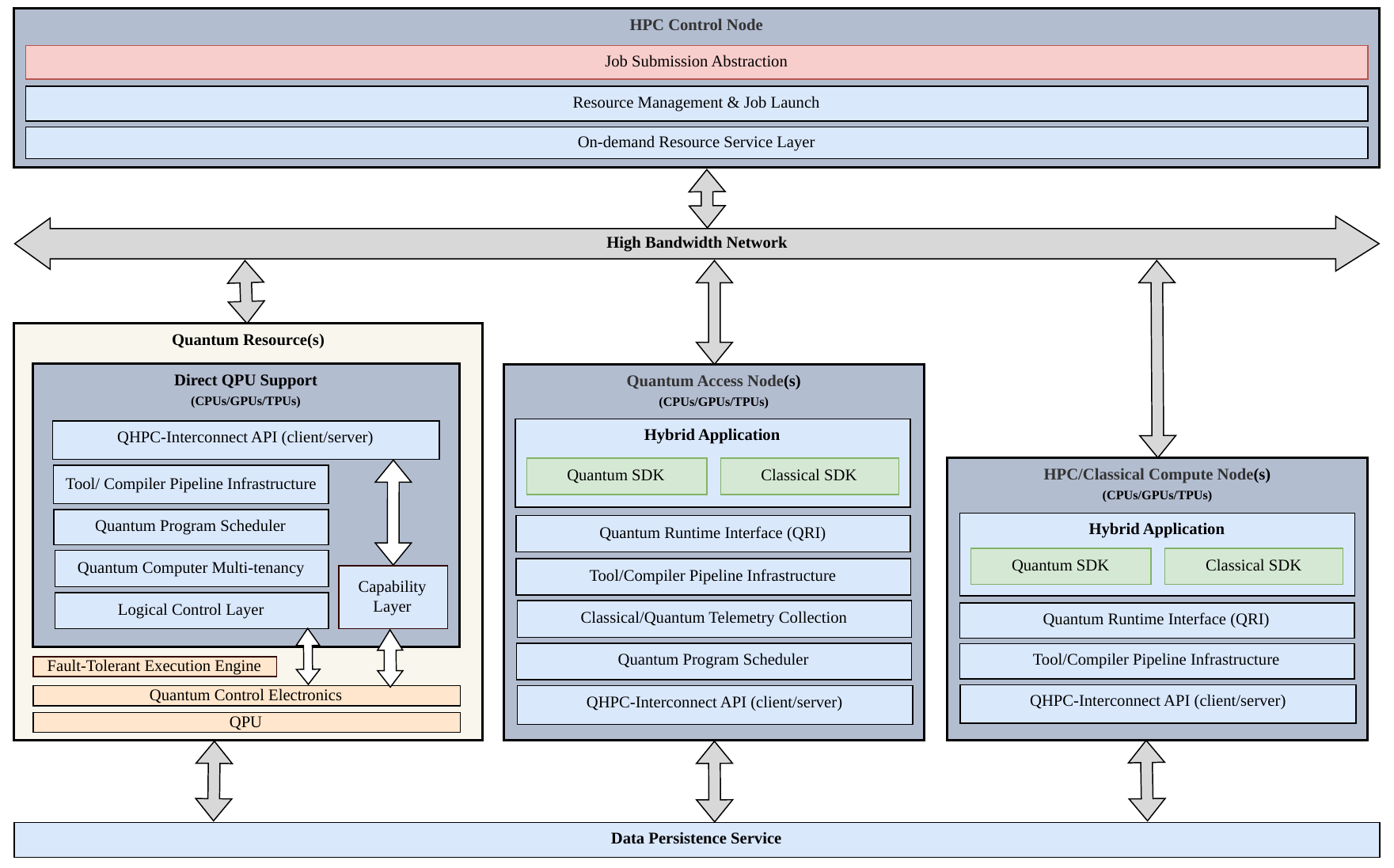}
      \caption{Proposed openQSE reference architecture for QHPC integration,
      showing a layered, deployment-agnostic stack with stable API contracts across
      control, access, and quantum-resource domains, including bidirectional QHPC
      interconnect, compiler/tool pipelines, scheduling/virtualization, and fault-
      tolerant execution paths. Not all layers may be needed
      in all deployments. For example, the Logical Control
      Layer would not be needed for a NISQ deployment.}
      \label{fig:openqse-architecture}
\end{figure*}

As shown in \autoref{fig:openqse-architecture}, the architecture is organized into five
logical parts. These are an \ac{HPC} control plane, one or more
\ac{HPC}/classical compute nodes, one or more quantum access nodes, one or
more quantum resources, and shared infrastructure for network and data persistence.
This decomposition illustrates a representative mapping of architectural
responsibilities onto the heterogeneous compute elements typically found in HPC
centers. While alternative mappings are possible, the configuration presented here
is chosen to clearly separate concerns and align functional responsibilities with
the capabilities of each component. Not all layers or compute elements are required
in every deployment; the figure presents a comprehensive view of the architecture,
with specific configurations determined by the target environment.

At the top of the stack, the \ac{HPC} control node provides the logical control
plane for job submission and resource abstraction. It may map to login nodes,
workflow services, or control microservices rather than a single fixed system. The
job submission abstraction layer is a standardized, system-agnostic representation of quantum resources and their capabilities. In practice, \acp{RMS} (e.g., SLURM or Flux)
each expose different mechanisms for describing and requesting resources; without a
common model, this leads to fragmentation and non-portable job specifications.
openQSE addresses this by defining a consistent resource description that decouples
job submission from any specific \ac{RMS}, enabling the same job specification to be
submitted across heterogeneous systems without modification. This also supports
deployments in which quantum resources reside in different administrative
domains. The on-demand resource service layer
supports policy-bound expansion, allowing workflows to request temporary additional
resources within declared burst envelopes.

The quantum resource defines the execution boundary for quantum workloads, spanning the QHPC-interconnect for quantum task submission down to physical pulse delivery.
This
boundary also includes the tool pipeline for resource-side lowering, the quantum program
scheduler and multi-tenancy layers, and a \ac{LCL} that mediates access to hardware
capabilities.

The scheduler and multi-tenancy layers manage how quantum workloads share a device while enforcing hardware constraints. In most current systems, only one circuit executes at a time, so these layers queue workloads according to defined policies. As hardware evolves to support concurrent execution across separate qubit regions, the same layers can be extended to schedule and manage parallel workloads.

The \ac{LCL} operates in \ac{RTD}. In the \ac{FTQC} era, \ac{RTD} represents the logical time
of quantum operations. In both the \ac{NISQ} and \ac{FTQC} eras this layer exposes functionality such as parameter streaming, other
device-specific features, such as incremental just-in-time (JIT) compilation, and support for hierarchical error correction codes. It
may also expose capability discovery features, which may be statically configured or
dynamically read from the device. The \ac{FTEE} operates in the \ac{DTD}. It
consumes \ac{PTD} execution state, makes bounded-
latency adaptive decisions, and either issues corrective actions or escalates
upward when the latency budget allows it. This interpretation aligns with
recent \ac{QEC} literature on active reset, conditional control, and dynamic
circuits~\cite{active-reset,conditional-op,dynamic-circuits}. As a
result, the degree of \ac{FTEE} exposure to higher layers may vary across
modalities. The architecture separates \ac{LCL} and \ac{FTEE} to make the intent
explicit.

The quantum access node is the operational entry point for many workflows. It
can host user-facing hybrid scripts or service-side preparation logic, and it
can also act as a richer cluster tier with direct access to one or more quantum
resources. \ac{HPC}/classical compute nodes share most of the same runtime
interfaces, while telemetry and shared operational services can be centralized
outside the main execution path. Data persistence services and high-bandwidth networking
support staging, classical checkpointing, and data movement across local and remote
resources.

\subsection{Quantum Runtime Interface}

The \ac{QRI} is the application-facing runtime boundary. It separates
application logic from vendor-specific and site-specific execution details in a
way that is similar in intent to HIP~\cite{hip-docs} for heterogeneous GPU environments.
Applications can continue to use the preferred SDKs, but they interact with a
common execution contract rather than a different backend-specific runtime for
each target.

In \ac{openQSE}, \ac{QRI} is available wherever hybrid applications run,
including \ac{HPC}/classical nodes and quantum access nodes. It supports
resource-aware submission, lifecycle management, and multi-resource launch at
an abstraction level above low-level resource interfaces. This reduces direct
application dependence on backend-specific interfaces and helps preserve
portability when deployment environments change.

At the same time, current quantum hardware exhibits significant device-specific
variability, and even the same system may change characteristics over time.
While this limits the degree of uniformity that \ac{QRI} can enforce, it remains a
useful abstraction that can accommodate such variability through extensible
interfaces and capability-aware execution models.

\subsection{QHPC-Interconnect API}

The \ac{QHPC}-Interconnect API is a bidirectional interface that is
the architectural boundary between different compute elements. It supports
downstream and upstream invocation through an RPC layer, which allows different APIs to be implemented seamlessly, such as submit, configure, execute, retrieve
operations, upstream signaling for events, callbacks,
adaptation requests, and escalation triggers.

This interface domain also includes connection management,
authentication, data compression and protection, and abstraction over the
underlying high-speed fabric. These capabilities belong together because
latency, trust, and transport behavior are coupled in real deployments.
Bidirectional behavior matters when quantum-side services need upstream action,
including additional classical processing, calibration notification or dynamic resource
requests.

\subsection{Compiler Tool Pipeline}

The Tool/Compiler Pipeline Infrastructure is a governed model for
combining algorithm-specific transformation tools and compiler passes in
ordered stages. Pipelines are bound to inputs and workflows, so that a given
path applies a consistent set of transformations. This supports
reproducibility, policy enforcement, and performance tuning while still
leaving room for domain-specific extensions.

Pipeline placement remains flexible. Some stages can run on
\ac{HPC}/classical nodes, while later backend-specific lowering can run on
access nodes or near the quantum resource. The architecture does not force one
compilation
location. It standardizes how stages are described, connected, and
validated. That makes it easier to reuse transformations across frameworks
and gives ecosystem efforts, such as Jeff~\cite{Jeff,jeff2}, a clearer architectural
target.

\subsection{Control Electronics}
\label{sec:control-electronics}

Control electronics sit at the boundary between the \ac{PTD} and the \ac{DTD}.
They translate bounded-latency digital commands into physical qubit operations,
including waveform generation, pulse modulation, readout, and real-time
feedback. Their behavior affects scheduling, throughput, and \ac{FTEE} design,
so they should be treated as an architectural concern rather than a purely
vendor-internal detail.

IonQ makes this boundary especially visible. In the Forte system, an embedded
controller applies RF pulse sequences for coherent gate execution,
while the runtime maintains pulse parameters through recurring calibration
cycles between circuit batches. Those cycles consume
scheduler-visible \ac{QPU} time. Reported measurements indicate that
calibration and testing consume about 47\% of total operational time, leaving
53\% for user circuit execution. IonQ addresses this with a
nanosecond-precision real-time tier and an upper-layer performance model that
uses native gate parameters and hardware state for scheduling and pipelining
\cite{openqse-workshops}. On IonQ Forte, single-qubit gates are reported at
110--130~$\mu$s, two-qubit  gates at 900--950~$\mu$s, and ion-chain
cooling at about 3~ms per shot \cite{ionq-forte}.

More generally, control electronics are moving beyond fixed waveform delivery.
Parameter streaming, conditional branching, and feedback paths are becoming
architecturally relevant capabilities. \ac{openQSE} therefore needs a
capability-discovery interface so upper layers can query supported timing
windows, branching primitives, and feedback latencies instead of assuming the
same path on every platform. That contract also needs to remain stable enough
for \ac{QEC}-driven corrective actions to be dispatched without redesigning the
surrounding software stack.

\section{Conclusion and Future Work}

This paper examined nine representative \ac{QHPC} software stacks along four classification axes to highlight areas of convergence as well as ongoing divergence. Current systems already offer useful building blocks for hybrid execution, but they differ in key aspects. These include how runtime boundaries are defined, how interconnect semantics is handled, what level of observability is provided, and what hardware capabilities are exposed to higher software layers. We proposed the \ac{openQSE} architecture, a composable layer model for operating across cloud and on-premises environments.

This work has three main limitations. First, the analysis is based on
publicly documented capabilities and vendor-described stack descriptions
as of April~2026; the \ac{QHPC} landscape is evolving rapidly and
specific stack characteristics may shift. Second, the \ac{openQSE}
architecture is presented as a conceptual reference model rather than
a reference implementation, and its concrete interface contracts
remain to be specified. Third, no cross-stack quantitative performance
comparison is provided; structural rather than measurement-based
characterization is the focus of this survey.

Future work will include a structured study to evaluate each stack by collecting a wide range of metrics, such as execution latency, throughput, resource utilization, and integration overhead. This will provide insight into the best implementation for specific features and reveal where designs may be improved. Developing the \ac{openQSE} reference implementation will also provide a practical environment to integrate components and expose interoperability issues early. This creates a feedback loop where architectural decisions are validated in code. Findings can then guide iterative refinement and be shared with the community through contributions to open-source projects.

\section*{Acknowledgments}

The authors acknowledge the contributions of IBM for their insights on QHPC software stack design, as well as their participation in discussions and feedback that helped shape the perspectives presented in this paper.

The authors used Claude (Anthropic), Gemini (Google), Microsoft Copilot, and ChatGPT and Codex (OpenAI) for editorial assistance. Final document was reviewed and verified by the authors, who are solely responsible for the technical content.

\bibliographystyle{IEEEtran}
\bibliography{references}

\end{document}